# Direct evidence of low work function on $SrVO_3$ cathode using thermionic electron emission microscopy and high-field ultraviolet photoemission spectroscopy

Md Sariful Sheikh [a], Lin Lin [a,1], Ryan Jacobs [a], Martin E. Kordesch [b], Jerzy T. Sadowski [c], Margaret Charpentier [d], Dane Morgan [a, *], John Booske [e, *]

[a] Department of Materials Science and Engineering, University of Wisconsin-Madison, Madison, WI, 53706, USA.

[b] Department of Physics and Astronomy, Ohio University, Athens, OH 45701, USA.

[c] Center for Functional Nanomaterials, Brookhaven National Laboratory, Upton, NY 11973, United States.

[d] Kimball Physics, Inc., Wilton, NH 03086, USA.

[e] Department of Electrical Engineering and Computer Science, University of Wisconsin-Madison, Madison, WI, 53706, USA.

Corresponding authors Email: ddmorgan@wisc.edu, jhbooske@wisc.edu

## Abstract:

Perovskite $SrVO_3$ has recently been proposed as a novel electron emission cathode material. Density functional theory (DFT) calculations suggest multiple low work function surfaces and recent experimental efforts have consistently demonstrated effective work functions of ~2.7 eV for polycrystalline samples, both results suggesting, but not directly confirming, some fraction of even lower work function surface is present. In this work, thermionic electron emission microscopy (ThEEM) and high-field ultraviolet photoemission spectroscopy are used to study the local work function distribution and measure the work function of a partially-oriented-(110)-$SrVO_3$ perovskite oxide cathode surface. Our results show direct evidence of low work function patches of about 2.1 eV on the cathode surface, with corresponding onset of observable thermionic emission at 750 °C. We hypothesize that, in our ThEEM experiments, the high applied electric field suppresses the patch field effect, enabling the direct measurement of local work functions. This measured work function of 2.1 eV is comparable to the previous DFT-calculated work function value of the SrVO-terminated (110) $SrVO_3$ surface (2.3 eV) and SrO terminated (100) surface (1.9 eV). The measured 2.1 eV value is also much lower than the work function for the

[1] Present address: Andlinger Center for Energy and the Environment, Princeton University, Princeton, NJ 08540.

(001) $LaB_6$ single crystal cathode (~2.7 eV) and comparable to the effective work function of B-type dispenser cathodes (~2.1 eV). If $SrVO_3$ thermionic emitters can be engineered to access domains of this low 2.1 eV work function, they have potential to significantly improve thermionic emitter-based technologies.

# 1 Introduction:

Electron emission cathodes are key components of various high frequency, high power vacuum electronic devices such as traveling wave tubes and klystrons, and lower power device applications such as electron microscopes, ion thrusters for satellites, and thermionic energy converters of concentrated sunlight and industrial waste heat [1-4]. The electron emission from the cathode is largely determined by which surface terminations and orientations are present on the emitter, and their respective work function values. A stable, low work function surface is desired to provide efficient electron emission current during cathode operation, where emission may be controlled by incident photons (photoemission), high temperature (thermionic emission), electric field (field emission), or some combination [5-8]. Over the past century, significant research efforts have been devoted to discovering and engineering new low work function cathode materials and reducing the work function of existing cathode materials [9-17].

The perovskite oxide $SrVO_3$ is a promising next-generation electron emission cathode due to its high electrical conductivity, ease of bulk synthesis, and low predicted work function [18-21]. The cubic $SrVO_3$ perovskite structure (space group Pm-3m, No. 221) consists of alternating polar SrO and $VO_2$ layers along the crystallographic [001] direction, and alternating SrVO and O layers along the [110] direction. According to density functional theory (DFT) calculations, the SrO-terminated (001) surface and SrVO-terminated (110), equivalent to (011), surfaces have low calculated work functions of just 1.9 eV and 2.3 eV, respectively. These low work function values are enabled by the presence of large intrinsic surface dipoles [18]. Experimental studies using thermionic emission current measurements of bulk polycrystalline $SrVO_3$ have confirmed that $SrVO_3$ has a reproducible effective work function (defined and discussed more below) of 2.7 – 2.8 eV. In a few cases, these experiments also revealed a quite low effective work function of ~2.3 eV from the bulk polycrystalline $SrVO_3$ cathode surface [19], suggesting potentially very low work functions can be obtained in this material. Furthermore, in a recent study, the bulk polycrystalline $SrVO_3$ cathode demonstrated stable emission during cyclic heating and cooling, and drift-free steady electron emission during one hour of a continuous thermionic emission test [22], further demonstrating the promising potential of the $SrVO_3$ cathode.

It is important to note that these previous experimental measurements of thermionic emission from the heterogeneous emitting surface of polycrystalline $SrVO_3$ were conducted using a relatively low applied electric field. The use of a low applied electric field produces measurements of "effective" work function, i.e., a complex average of many surface terminations, as opposed to directly revealing the lowest local work function. The surface electronic physics producing an effective work function under low applied field is termed the patch field effect [14]. According to the patch field effect, the higher work function patches surrounding the low work function patch pull the emitted electrons back towards the surface, resulting in a higher energy barrier for electron emission from the low work function surface patches than expected based on their local work function values. This produces an effective work function for the heterogenous cathode surface that is higher than the lowest local work function surface present. Recent studies have unveiled the role of patch fields in physical models of cathode surfaces as well as actual surfaces from commercial cathode specimens [23-26]. Previous experiments reported the effective thermionic work function of bulk polycrystalline $SrVO_3$ cathode surface measured with a low applied electric field of $\sim 1\times 10^6$ V/m. For these surfaces, patch field analysis suggested that a field of at least several times of $10^6$ V/m would be required to cancel the patch fields and reveal the lowest local work functions of $SrVO_3$ [19].

In this work, we studied the thermionic emission behavior and work function of (110)-$SrVO_3$ cathode surfaces using the thermionic electron emission microscopy (ThEEM) imaging technique. The experimental study revealed the presence of other crystallographic orientations along with the main (110) orientation in the crystalline specimen, prompting us to refer to this specimen as a *partial-(110)-$SrVO_3$* cathode surface. The use of ThEEM measurements is advantageous for understanding local emission on cathode surfaces because it is performed by applying a large electric field ($\sim 10^7$ V/m) to the cathode surface. Such large electric fields can cancel the patch field effect and enable local work function measurements of isolated $SrVO_3$ domains. We also studied the distribution of the local work function patches on the partial-(110)-$SrVO_3$ cathode surface.

The work function measurement at 950 $^{o}$C revealed a value as low as 2.1 eV, likely within the uncertainty of the DFT-calculated work function of 2.3 eV for the SrVO-terminated (110) plane and 1.9 eV of SrO terminated (100) plane of cubic $SrVO_3$ [15, 18]. This 2.1 eV work function value of (110) $SrVO_3$ cathode is even lower than the previously observed lowest effective work function value of $SrVO_3$ polycrystalline cathode (~2.3 eV). However, this finding is consistent with the possibility that some fraction of the surface of these previously examined polycrystalline $SrVO_3$ samples had a work function of ~2.1 eV. The obtained work function is significantly lower than the (001) $LaB_6$ single crystal cathode work function of ~2.7 eV [27] and is comparable to the effective work function of a B-type dispenser cathode (~2.1 eV) [28], suggesting that $SrVO_3$ is a promising thermionic emission material.

## 2 Experimental details:

The material examined in this study was a previously synthesized $SrVO_3$ crystalline specimen obtained from Kimball Physics Inc. The detailed synthesis process of the $SrVO_3$ crystalline specimen and its characterizations are reported in previous work [29], and we provide a summary here. The $SrVO_3$ crystalline specimen was prepared from a polycrystalline mixture of $SrVO_3$ (85 %) and $Sr_2V_2O_7$ (15 %) using a laser floating zone growth technique. At first, the isostatically pressed $SrVO_3/Sr_2V_2O_7$ cylindrical rod of diameter 4 mm and length 80 mm was sintered at 1200 $^{o}$C for 8 h under vacuum (~$10^{-5}$ torr). The crystalline specimen was prepared at 1 atm pressure under Ar:$H_2$ (95%:5%) gas flow in the TiltLDFZ furnace (Crystal Systems Inc.) equipped with 5 × 200 W GaAs lasers. During the synthesis, the top part of the seed rod placed in the molten zone was melted and then the feed rod was brought into contact with the seed rod in the molten zone. During the growth, both the rods were counter-rotated, and the crystalline specimen was prepared by moving both the rods downward while keeping the molten zone fixed at the same position. Laue diffraction revealed a [110] direction along the length of the cylindrical crystal as reported before [29]. However, using electron backscatter diffraction (EBSD) at two separate positions on the crystal surface, we observed different Kikuchi patterns, suggesting the presence of other orientations along with the [110] orientation as discussed later. Based on the characterizations performed, including thermionic electron emission maps and EBSD, the evidence supports the conclusion that the specimen surface consists of an array of micro-patches with varying orientations, some of which are low work function facets, like (100)- or (110)- orientations. Henceforth, in the remainder of this paper, we refer to such a surface as a *partial-(110)-$SrVO_3$ surface.* To prepare the cathode, a part of the crystalline specimen was cut along the circular cross-section of the crystalline specimen to get the (110) surface and polished for the electron emission test.

Electron backscatter diffraction (EBSD) measurements were performed using a Tescan Vega3 scanning electron microscope at Kimball Physics Inc. Field emission electron microscopy (FESEM) images were collected using a Zeiss 1530 scanning electron microscope (SEM). Surface elemental stoichiometry was studied using an energy dispersive spectroscopy (EDS) function integrated with the Zeiss 1530 FESEM instrument.

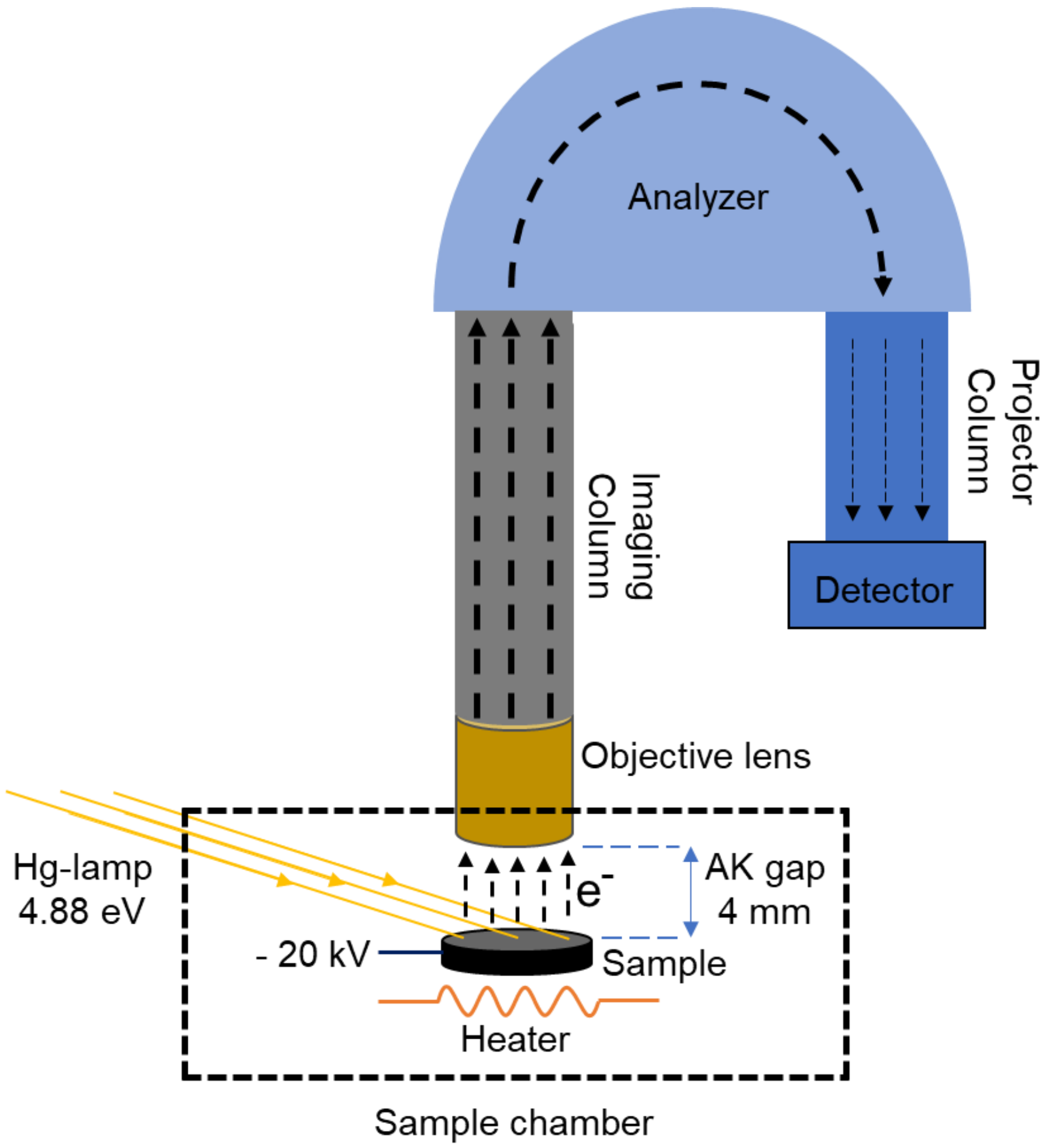

**Figure 1:** Schematic of ThEEM and UPS measurement configuration.

The ThEEM image and the thermionic emission measurements were performed using an Elmitec LEEM III microscope at the XPEEM/LEEM end station of the ESM beamline (21-ID) of the National Synchrotron Light Source II (NSLS-II) at Brookhaven National Laboratory (BNL). A simplified schematic of the ThEEM and work function measurement set up is shown in Figure 1. Before the electron emission test, the cathode was heated at 1300 $^{o}$C for 1 hour in ultra-high vacuum (base pressure of approximately $5 \times 10^{-10}$ Torr) to remove any surface contamination which may have occurred during air exposure and recover the surface from any damage caused by polishing. After heating, the cathode was cooled slowly to room temperature, and then it was reheated to high temperature for the ThEEM measurements at 750, 800, 850, 900 and 950 $^{o}$C. The work function of the cathode surface was measured at high temperatures of 900 and 950 $^{o}$C by ultraviolet photoemission spectroscopy (UPS) using a Hg lamp (4.88 eV) as the photon source. The temperature of the cathode was measured using a thermocouple integrated into the sample holder. During ThEEM and the UPS measurements, an applied electric bias voltage of -20 kV was applied at the cathode, and a cathode-surface-to-anode gap (AK gap) of ~ 4 mm was maintained. This applied electric field was nearly 5 times stronger than the one used in previously reported work function measurements of polycrystalline $SrVO_3$ cathodes [19]. The intensity map of the spatially nonuniform emission current in the ThEEM images was recorded in arbitrary units using the ImageJ software. The average intensity at different temperatures was measured from five different emission patches' intensity on the ThEEM images.

# 3 Results and Discussion:

ThEEM is an important research tool to study the local distribution of electron emission from a heterogeneous thermionic cathode surface. It also allows us to observe the emission at high temperatures under a high applied electric field, which cancels out the patch-field effect. Figures 2(a-e) show the ThEEM images of a 50 μm × 50 μm region on the cathode surface collected at 750, 800, 850, 900 and 950 $^{o}$C, respectively. The ThEEM images show that the measurable emission starts at around 750 $^{o}$C, suggesting the presence of low work function domains on the cathode surface. At this low temperature, we observe the electron emission is limited to a few small, low work function patches. However, as the temperature increased, more patches of progressively higher work function became activated and contributed to the overall emission. The patchy emission suggests the presence of non-uniform crystallographic grain orientations and/or terminations on the cathode surface. The emission current density trend, qualitatively indicated by the brightness of the ThEEM images, was observed to increase exponentially with temperature. Figure 2(f) shows the temperature-dependent trend of the ThEEM-image-averaged intensity of the electron emitting patches. This image-averaged intensity trend is qualitatively consistent with the Richardson-Dushman equation of thermionic emission current. Only a small fraction of the cathode surface is showing emission with low work function, indicating this $SrVO_3$ cathode surface is a very patchy emitter, similar to observations made on previous polycrystalline specimens [19, 22].

We also used ThEEM to study the emission from a polycrystalline $SrVO_3$ cathode surface at a higher temperature of 1100 $^{o}C$ as shown in Figure S2. ThEEM measurements of the polycrystalline sample reveal the presence of bright emission spots on the cathode surface very similar to those observed in Figure 2. The bright emission spots, which are the low work function regions, could originate from the low work function (110)-oriented SrVO-termination and (100)-oriented SrO-termination surfaces discovered using the DFT calculations. However, from this study we cannot definitively identify the orientation or termination of the observed low work function regions, and doing so would require more detailed fundamental characterizations of the material.

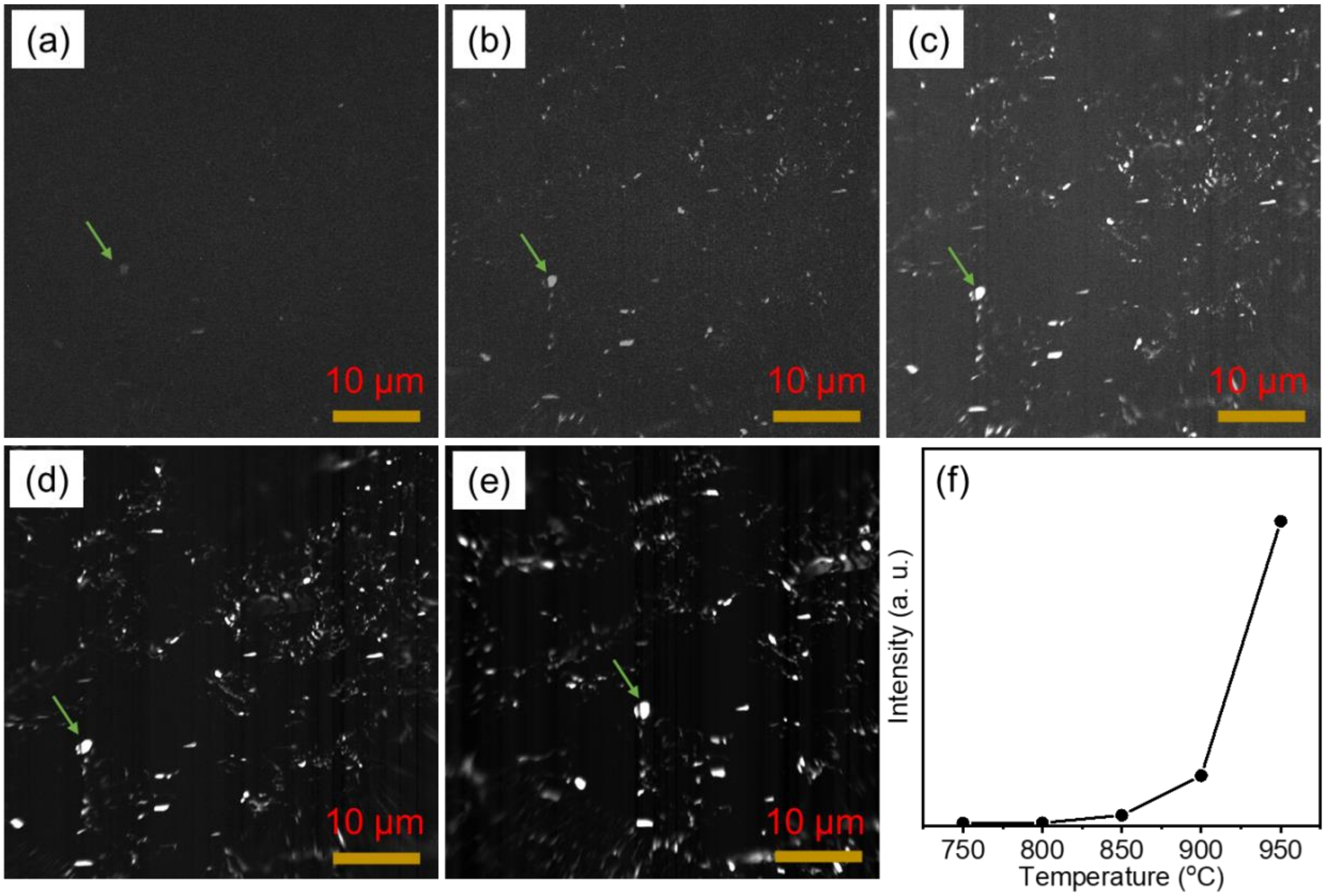

**Figure 2:** ThEEM image of the (110) $SrVO_3$ cathode surface collected at (a) 750 °C, (b) 800 °C, (c) 850 °C, (d) 900 °C and (e) 950 °C. (f) Temperature dependence of the image-averaged intensity of the emitting patches as observed in the ThEEM images. The green arrows in (a)-(e) are a guide for the eye, marking the same emission spot in each image.

The kinetic energy distribution of the thermally and photothermally emitted electrons was measured using a LEEM III energy analysis system at BNL. The LEEM microscope was operated at an accelerating voltage of 20 kV. The emitted electrons from the cathode passed through the hemispherical electron analyzer system as shown in Figure 1, and generated energy filtered images at varying kinetic energies [30]. Figures 3(a-b) show pure thermionic emission current versus kinetic energy data for the cathode surface at 900 and 950 °C, respectively. The measured thermionic emission peak shifts toward lower kinetic energy as the temperature increases, suggesting a decrease in effective work function with increasing temperature. The work function of the cathode was measured using ultraviolet photoemission spectroscopy (UPS) (Hg-lamp, 254 nm, energy threshold 4.88 eV) at temperatures 900 and 950 °C. Figures 3(c, d) show the kinetic energy distributions of the thermal- and photo-generated electrons from the cathode surface at 900 and 950 °C, respectively. The UPS valence band spectra do not show any shift in the valence band maximum position (A) as observed in Figures 3(c, d). Meanwhile, the shifting of the secondary electron cutoff energy position (B) towards lower kinetic energy at 950 °C as compared to 900 °C suggests a reduction of the effective work function at 950 °C [14, 31]. In UPS, the kinetic energy (KE) of an emitted electron is $KE = h\nu - \phi$, where h is Planck's constant, ν is the incident photon frequency and ϕ is the work function. Hence, the work function is measured by subtracting the energy difference between the thermionic electron emission peak maximum and the Hg lamp emission data tail (high kinetic energy cutoff point) from the Hg lamp photon energy (4.88 eV) [14, 30, 31]. The high applied electric field in ThEEM is expected to cancel the patch field effect from the high work function region surrounding the low work function patches (observed in Figure 2), enabling us to measure the local work function of the low work function patches [14]. The measured work function values at temperatures 900 °C and 950 °C are 2.47 and 2.10 eV, respectively. The measurement shows a decreased work function at 950 °C as compared to the work function measured at 900 °C. A temperature-dependent shifting of the work function has also been observed in other patchy emitters, including $Sc_2O_3$-BaO-coated-W and polycrystalline $SrVO_3$ cathodes [19, 30]. We speculate that the surface domains which have bulk orientations consistent with lowest work function reach the required surface termination or surface structure at around 950 °C and start emitting detectable electron current. However, understanding the origin of this work function change requires more in-situ research on the surface structure-work function correlation of the $SrVO_3$ cathode. The measured 2.1 eV work function is comparable with the DFT-calculated 2.3 eV work function of the SrVO-terminated [110] surface and 1.9 eV work function of the SrO-terminated (100) $SrVO_3$ surface [18]. To the best of our knowledge, this 2.1 eV effective work function at 950 °C is the lowest experimentally measured local work function reported from any bulk conductive oxide material not relying on volatile surface species or doping to produce the low work function.

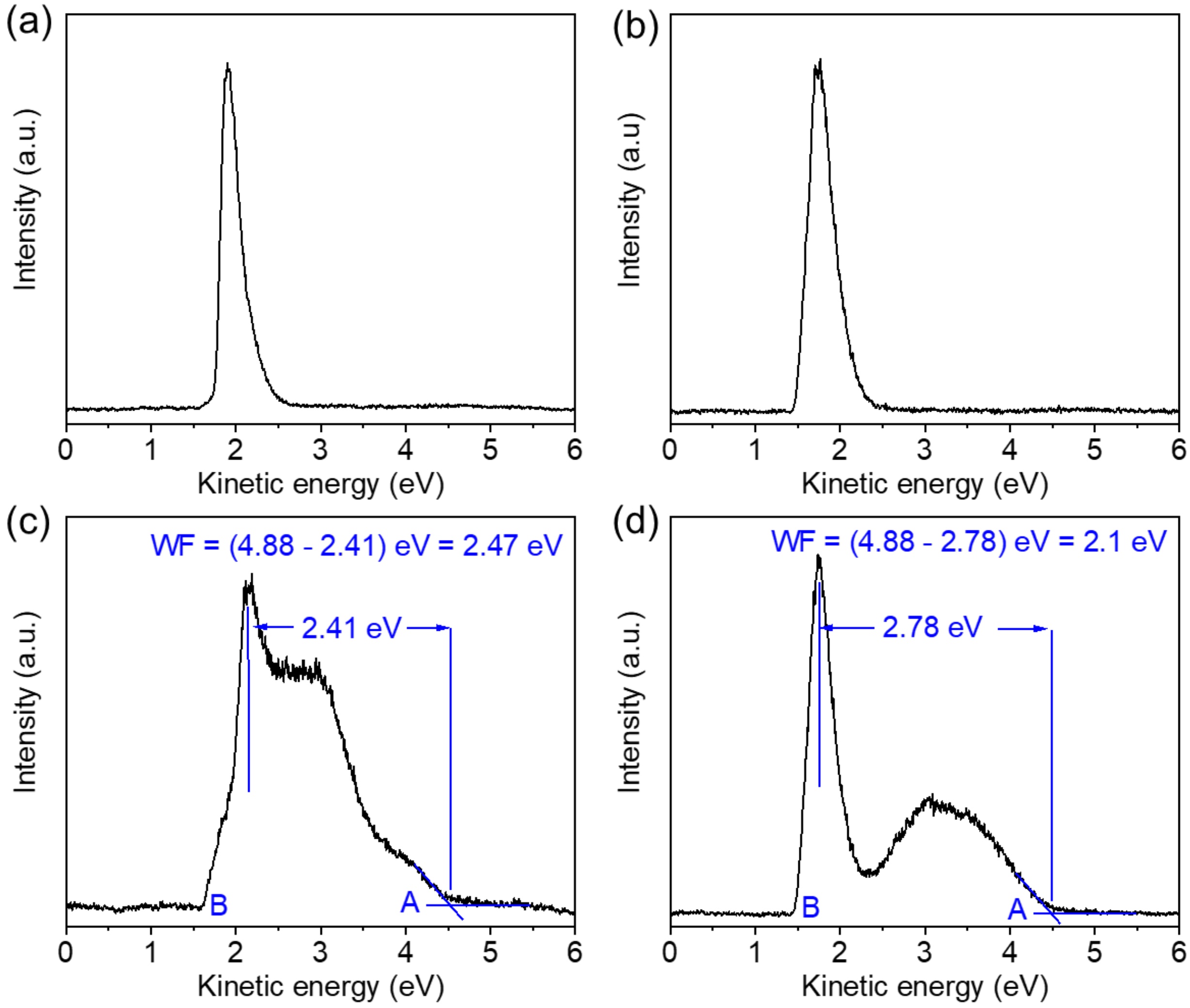

**Figure 3:** Thermionic and photoemission current measurements of the heated $SrVO_3$ cathode. Purely thermionic electron emission peak of the $SrVO_3$ cathode (without photoemission) at temperatures (a) 900 °C and (b) 950 °C. Combined thermionic and ultraviolet photoemission spectroscopy (UPS) measurements using a Hg lamp (4.88 eV) at temperatures (c) 900 °C and (d) 950 °C.

A previous study confirmed the phase purity of the $SrVO_3$ crystal and stoichiometric elemental ratios [29]. Hence, the heterogeneity in the emission behavior of the $SrVO_3$ cathode is hypothesized to arise due to the presence of different crystallographic orientations and terminations on the emission tested cathode surface. To check the crystallinity of the $SrVO_3$ cathode, EBSD was performed on the polished cathode surface. The EBSD study as shown in Figures 4(a, b) reveals different patterns at two different positions on the sample surface. This result suggests the emitting surface has additional crystallographic domains other than the dominant (110) orientation. This heterogeneity is consistent with the ThEEM results in Figure 2 showing heterogeneous emission, consistent with some domains on the cathode surface having higher work function compared to the initially-activated 2.1 eV low work function patches.

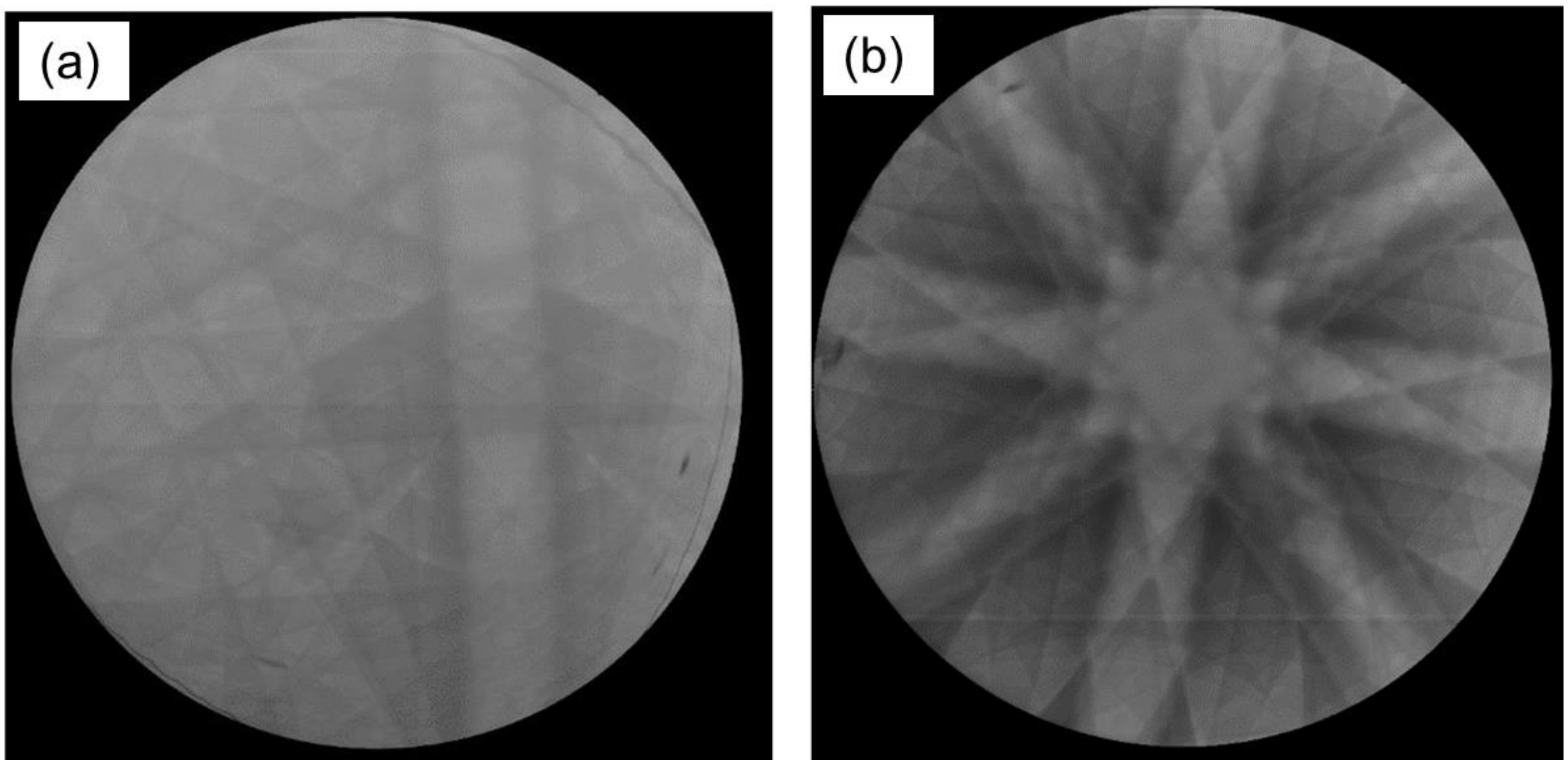

**Figure 4:** (a, b) SEM EBSD Kikuchi pattern at two different positions of the cathode surface.

To better understand the origin of the emission heterogeneity, we studied the elemental distribution on the cathode surface using EDS elemental mapping as shown in Figure 5. The EDS mapping shows a uniform distribution of the Sr, V and O atoms on the polished cathode surface. This data implies that the reason for the heterogeneity in the local work function on the cathode surface as observed in Figure 2 is not from major variations in the surface stoichiometry. However, the heterogeneity in the local work function may arise due to the presence of different crystallographic domains and terminations, consistent with the EBSD results in Figure 4, and only a small fraction of the cathode surface may have the (110)-oriented SrVO-termination and/or (100)-oriented SrO-termination, with the remainder of the surface comprising other surface orientations or terminations.

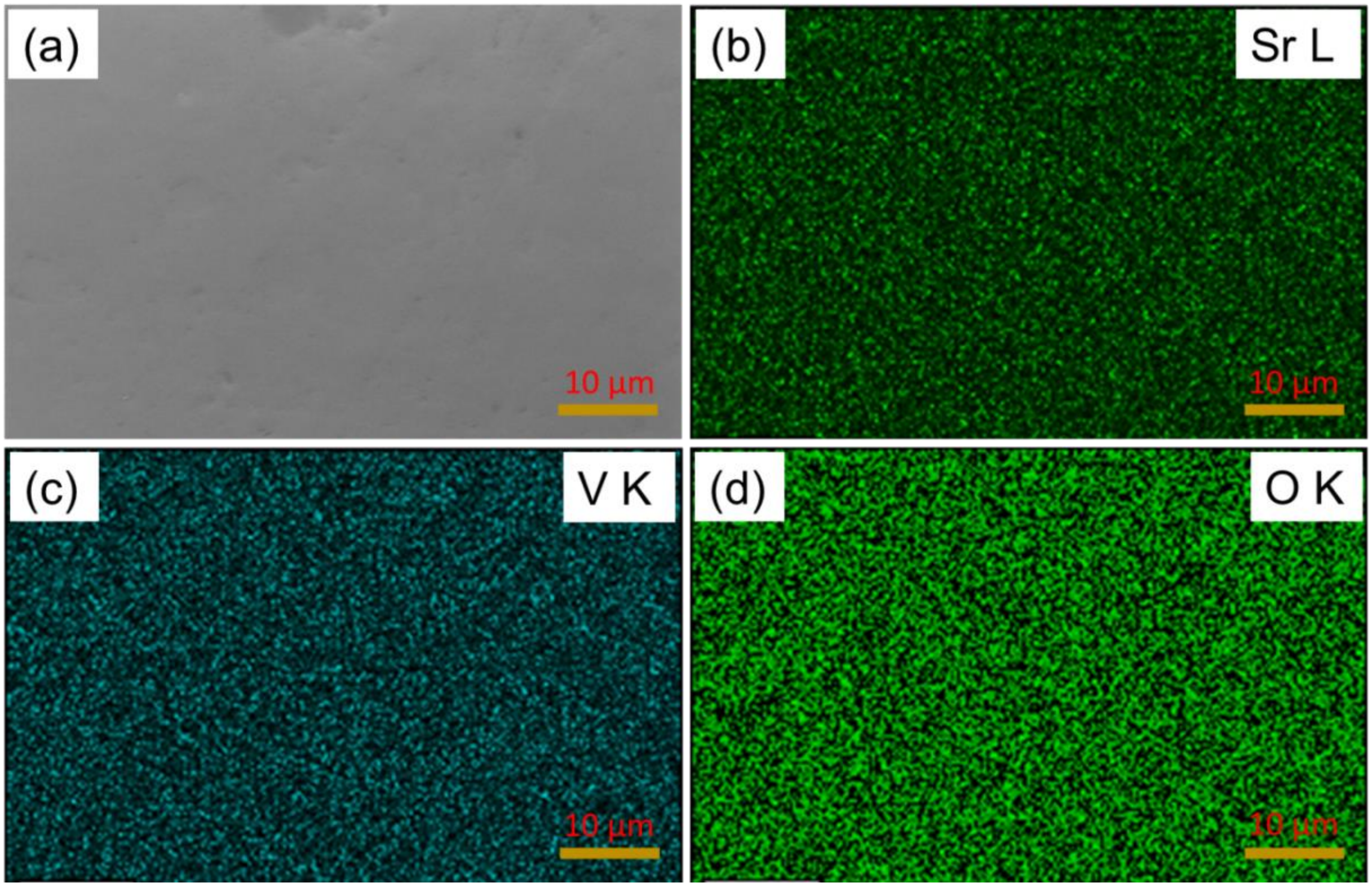

**Figure 5:** (a) FESEM image of $SrVO_3$ cathode surface. (b) Sr, (c) V and (d) O EDS elemental mapping on the polished $SrVO_3$ cathode surface. The scale bar is 10 µm in all figures.

A key challenge for future work is to identify the crystallographic orientation and termination of the surfaces giving rise to the low work function and then seek to increase their role in the emission. If the areas of the low work function patches were increased, or these areas occupied a larger fraction of the emission surface in the form of a higher density of small, copiously emitting patches, it could provide high emission current at reduced temperature and applied electric field (i.e., the field required to cancel the patch effects would be reduced). Such surface engineering would provide a promising path toward realizing $SrVO_3$ cathodes with enhanced emission, increasing the value of $SrVO_3$ emitters for a range of applications.

# 4 Conclusions:

Conducting perovskite oxide $SrVO_3$ is a promising thermionic cathode material due to its low effective work function. In this study, we used thermionic electron emission microscopy to study the local work function distribution on the partial-(110)-$SrVO_3$ cathode surface. ThEEM demonstrated the presence of low work function patches on the surface which started emitting at around 750 °C. The presence of low work function patches on the partial-(110)-$SrVO_3$ surface was revealed by using a high applied electric field to cancel the patch field effect, where these local work function patches had a low work function value of just 2.1 eV, much lower than (001) $LaB_6$ single crystal cathodes and comparable to B-type dispenser cathodes. Overall, this study provides two important results: (1) experimental verification that a high applied surface electric field enables emission-based local measurements of low work function patches on heterogeneous surfaces, and (2) verification of the existence of patches on a heterogeneous $SrVO_3$ surface that have work functions comparable to the DFT-predicted values of 2.3 eV for a SrVO-terminated (110) surface and 1.9 eV of SrO-terminated (100) surface of cubic $SrVO_3$, further illustrating the potential of $SrVO_3$ as a thermionic emission material.

## 5 Acknowledgements:

The authors would like to thank the Wisconsin Alumni Research Foundation (WARF) for providing financial support for this work. This research used resources of the Center for Functional Nanomaterials and the National Synchrotron Light Source II, which are U.S. Department of Energy (DOE) Office of Science facilities at Brookhaven National Laboratory, under Contract No. DE-SC0012704.

## 6 Authors contribution:

Md S. Sheikh: Investigations, discussion, writing (first draft), review and editing; L. Lin: Investigations, discussion, review, and editing; R. Jacobs: Conceptualization, funding acquisition, discussion, review and editing; Martin E. Kordesch: Investigations, discussion, review, and editing; Jerzy T. Sadowski: Investigations, discussion, review, and editing; Margaret Charpentier: Investigations and discussion; Dane Morgan: Conceptualization, funding acquisition, discussion, review and editing; J. Booske: Conceptualization, funding acquisition, discussion, review and editing.

## 7 Notes:

The authors declare no competing financial interest.

# 8 Data availability statement:

The raw ThEEM, UPS, EBSD, FESEM and EDS data as reported in Figures 2 through 5 are openly available in Figshare at https://doi.org/10.6084/m9.figshare.25202624 [32].

# Supporting Information for:

## Direct evidence of low work function on $SrVO_3$ cathode using thermionic electron emission microscopy and high-field ultraviolet photoemission spectroscopy

Md Sariful Sheikh [a], Lin Lin [a,1], Ryan Jacobs [a], Martin E. Kordesch [b], Jerzy T. Sadowski [c], Margaret Charpentier [d], Dane Morgan [a, *], John Booske [e, *]

[a] Department of Materials Science and Engineering, University of Wisconsin-Madison, Madison, WI, 53706, USA.

[b] Department of Physics and Astronomy, Ohio University, Athens, OH 45701, USA.

[c] Center for Functional Nanomaterials, Brookhaven National Laboratory, Upton, NY 11973, United States.

[d] Kimball Physics, Inc., Wilton, NH 03086, USA.

[e] Department of Electrical Engineering and Computer Science, University of Wisconsin-Madison, Madison, WI, 53706, USA.

Corresponding authors Email: ddmorgan@wisc.edu, jhbooske@wisc.edu

## S1. Synthesis of polycrystalline $SrVO_3$ cathode:

To prepare the cathode, first, the $SrVO_3$ perovskite oxide powder was synthesized using the sol-gel method as reported in previous report [19, 22]. For 1 gm $SrVO_3$ powder synthesis, stoichiometric amount of $Sr(NO_3)_2$ (Sigma Aldrich, 99.9%), $NH_4VO_3$ (Sigma Aldrich, 99%) and 25 gm citric acid (Dot Scientific Inc.) were dissolved in 250 ml of deionized water. The solution was then heated on a hot plate at 80 $^oC$ and continuously stirred using a magnetic stirrer until most of the water is evaporated and a bluish gel is formed. The gel was then heated at 120 $^oC$ for 12 hours in a box furnace. After drying the gel at 120 $^oC$, the furnace temperature was slowly raised to 600 $^oC$ (heating rate ~ 2 $^oC$/min) and sintered in stagnant air at 600 $^oC$ for 12 hours. Finally, the ground off-white precursor powder was calcined in a reduced environment at 1050 $^oC$ for 10 hours using a tube furnace and perovskite oxide $SrVO_3$ powder was synthesized. The reducing gas environment was maintained by continuously flowing a mixture of 5 % $H_2$ and 95 % Ar gas at a total flow rate of 200 SCCM.

The synthesized $SrVO_3$ powder was then pressed into cylindrical pellets of diameter ~ 3 mm and thickness ~ 0.75 - 1 mm using 300 MPa pressure in hydraulic press. Two of the compressed pellets were placed on each other and sintered at 1475 °C for 8 hours in a reducing environment maintained by flowing a gas mixture of 50 SCCM Ar and 10 SCCM of 5% $H_2$ balanced nitrogen gas. The heating and cooling rate were maintained at 5 °C /min. The synthesised pellet was polished on both sides before structural characterization and electron emission test. The X-ray diffraction (XRD) of cathode was studied using a Cu-Kα X-ray diffractometer (Bruker D8 Discovery). The XRD pattern as shown in Figure S1 confirms the cubic perovskite oxide phase formation in $SrVO_3$ cathode.

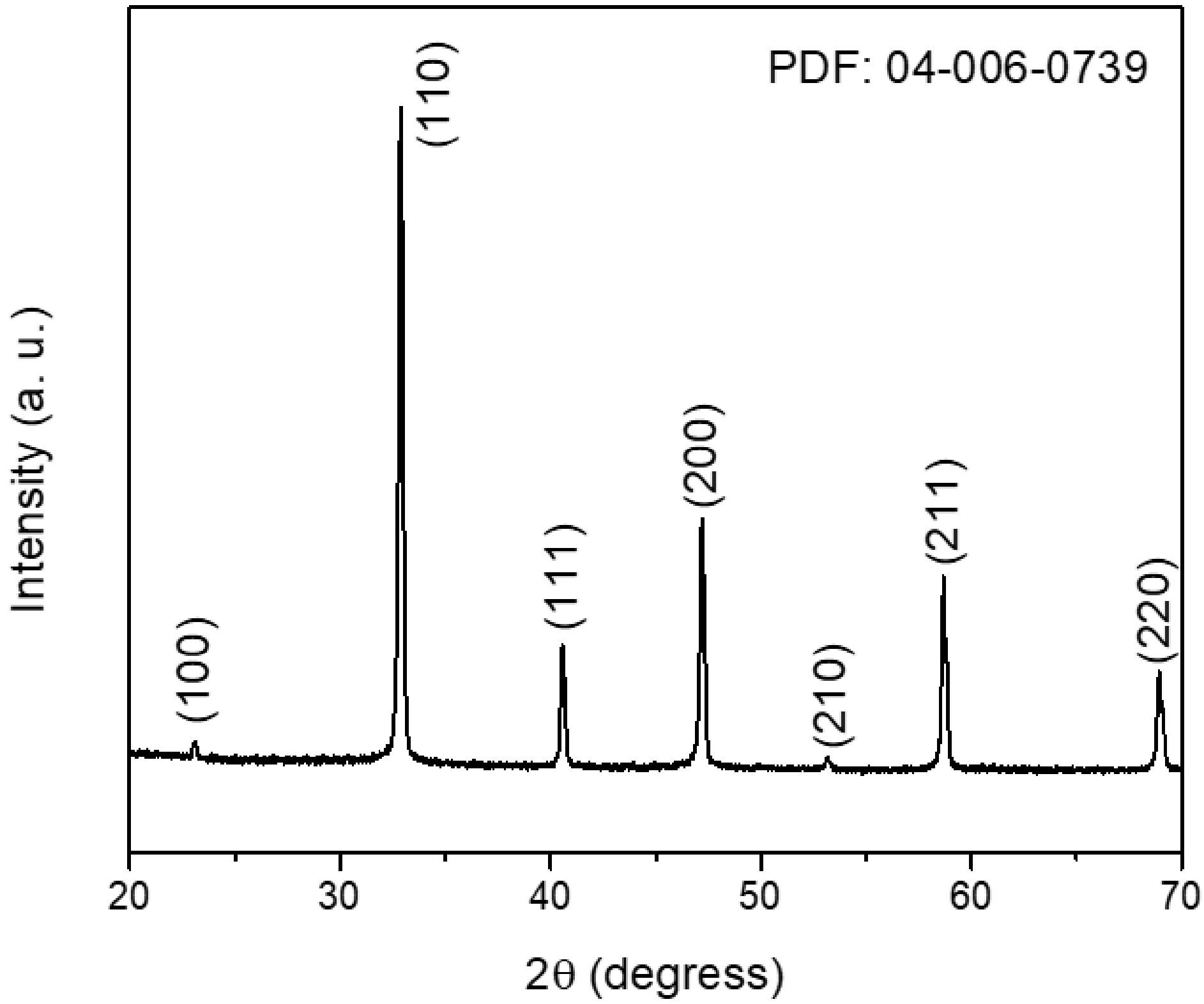

**Figure S1:** XRD pattern of the sintered $SrVO_3$ pellet surface confirming its polycrystalline nature and phase pure synthesis.

## S2. Electron emission test on polycrystalline $SrVO_3$ cathode:

The thermionic electron emission microscopy (ThEEM) measurement on the synthesized and polished polycrystalline $SrVO_3$ pellet was performed using ThEEM-PEEM based on the Bauer Telieps LEEM at the Ohio University [33]. During ThEEM measurement a 10 kV voltage difference was maintained between the cathode and anode separated at 4 mm, and the measurement was performed at 1100 °C.

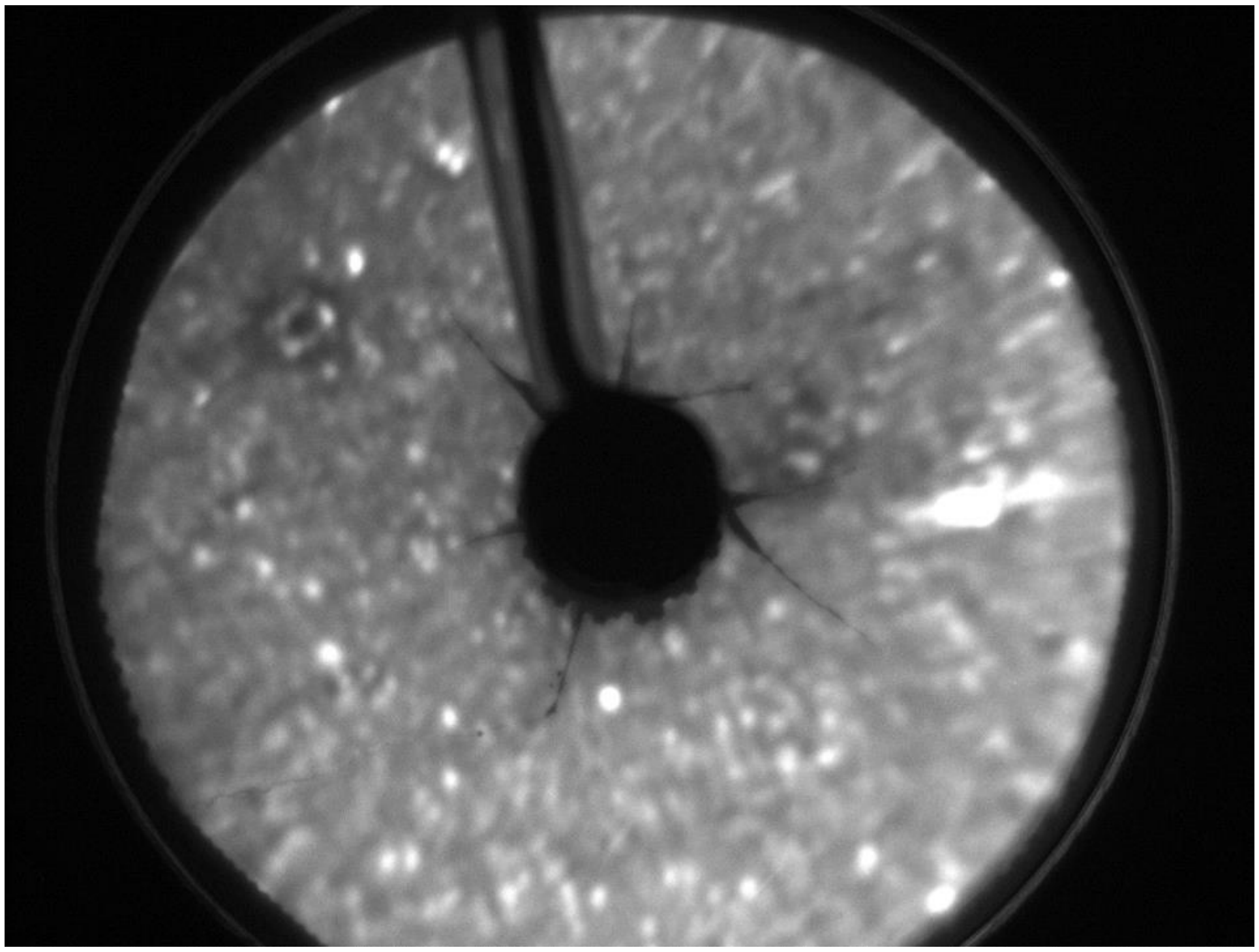

**Figure S2:** ThEEM image of the polycrystalline $SrVO_3$ cathode collected at 1100 $^{o}C$. The field of view is 400 μm.

## References:

[1] J. Y. Gao, Y. F. Yang, X. K. Zhang, S. L. Li, and J. S. Wang, "A review on recent progress of thermionic cathode," *Tungsten* 2, 289-300 (2020). https://doi.org/10.1007/s42864-020-00059-1

[2] G. Gaertner, and D. den. Engelsen, "Hundred years anniversary of the oxide cathode-A historical review," *Appl. Surf. Sci.* 251, 24-30 (2005). https://doi.org/10.1016/j.apsusc.2005.03.214

[3] K. Holste, P. Dietz, S. Scharmann, K. Keil, T. Henning, D. Zschätzsch, M. Reitemeyer, B. Nauschütt, F. Kiefer, F. Kunze, J. Zorn, C. Heiliger, N. Joshi, U. Probst, R. Thüringer, C. Volkmar, D. Packan, S. Peterschmitt, K. -T. Brinkmann, H.-G. Zaunick, M. H. Thoma, M. Kretschmer, H. J. Leiter, S. Schippers, K. Hannemann, and P. J. Klar, "Ion thrusters for electric propulsion: Scientific issues developing a niche technology into a game changer," *Rev. Sci. Instrum.* 91 (6), 061101 (2020). https://doi.org/10.1063/5.0010134

[4] J. Schwede, I. Bargatin, D. C. Riley, B. E. Hardin, S. J. Rosenthal, Y. Sun, F. Schmitt, P. Pianetta, R. T. Howe, Z.-X. Shen, and N. A. Melosh, "Photon-enhanced thermionic emission for solar concentrator systems," *Nature Mater.* 9, 762-767 (2010). https://doi.org/10.1038/nmat2814

[5] A. Kachwala, P. Saha, P. Bhattacharyya, E. Montgomery, O. Chubenko, and S. Karkare, "Demonstration of thermal limit mean transverse energy from cesium antimonide photocathodes," *Appl. Phys. Lett.* 123 (4), 044106 (2023). https://doi.org/10.1063/5.0159924

[6] F. Jin, and A. Beaver, "High thermionic emission from barium strontium oxide functionalized carbon nanotubes thin film surface," *Appl. Phys. Lett.* 110 (21), 213109 (2017). https://doi.org/10.1063/1.4984216

[7] L. Jin, Y. Zhou, and P. Zhang, "Direct density modulation of photo-assisted field emission from an RF cold cathode," *J. Appl. Phys.* 134 (7), 074904 (2023). https://doi.org/10.1063/5.0156328

[8] R. Ramachandran, and D. Biswas, "A unified thermal-field emission theory for metallic nanotips," *J. Appl. Phys.* 134, 214304 (2023). https://doi.org/10.1063/5.0173728

[9] S. Kimura, H. Yoshida, H. Miyazaki, T. Fujimoto, and A. Ogino, "Enhancement of thermionic emission and conversion characteristics using polarization- and band-engineered n-type AlGaN/GaN cathodes," *J. Vac. Sci. Technol. B* 39, 062207 (2021). https://doi.org/10.1116/6.0001357

[10] H. Yamaguchi, R. Yusa, G. Wang, M. T. Pettes, F. Liu, Y. Tsuda, A. Yoshigoe, T. Abukawa, N. A. Moody, and S. Ogawa, "Work function lowering of $LaB_6$ by monolayer hexagonal boron nitride coating for improved photo- and thermionic-cathodes," *Appl. Phys. Lett.* 122 (14), 141901 (2023). https://doi.org/10.1063/5.0142591

[11] M. M. Hasan, E. Kisi, and H. Sugo, "Structural and thermionic emission investigations of perovskite $BaHfO_3$ based low work function emitters" *Mater. Sci. Eng. B* 296, 116679 (2023). https://doi.org/10.1016/j.mseb.2023.116679

[12] S. Mondal, A. V. Rau, K. Lu, J.-F. Li, and D. Viehland, "Multicomponent hexaborides with low work functions by ultra-fast high temperature sintering," *Open Ceram.* 16, 100479 (2023). https://doi.org/10.1016/j.oceram.2023.100479

[13] T. Ma, R. Jacobs, J. Booske, and D. Morgan, "Discovery and engineering of low work function perovskite materials," *J. Mater. Chem. C* 9, 12778-12790 (2021). https://doi.org/10.1039/D1TC01286J

[14] L. Lin, R. Jacobs, T. Ma, D. Chen, J. Booske, and D. Morgan, "Work function: fundamentals, measurement, calculation, engineering, and applications," *Phys. Rev. Appl.* 19 (3), 037001 (2023). https://doi.org/10.1103/PhysRevApplied.19.037001

[15] T. Ma, R. Jacobs, J. Booske, and D. Morgan, "Understanding the interplay of surface structure and work function in oxides: A case study on $SrTiO_3$," *APL Mater.* 8, 071110 (2020). https://doi.org/10.1063/1.5143325

[16] L. Bai, T. Li, C. Zhang, H. Zhang, S. Yang, Q. Li, and Q. Sun, "Enhancing the thermionic electron emission performance of hafnium with nanocluster doping," *Appl. Phys. Lett.* 121, 061603 (2022). https://doi.org/10.1063/5.0106790

[17] T. Ma, R. Jacobs, J. Booske, and D. Morgan, "Work Function Trends and New Low-Work-Function Boride and Nitride Materials for Electron Emission Applications," *J. Phys. Chem. C* 125 (31), 17400-17410 (2021). https://doi.org/10.1021/acs.jpcc.1c04289

[18] R. Jacobs, J. Booske, and D. Morgan, "Understanding and controlling the work function of perovskite oxides using density functional theory," *Adv. Funct. Mater.* 26 (30), 5471-5482 (2016). https://doi.org/10.1002/adfm.201600243

[19] L. Lin, R. Jacobs, D. Chen, V. Vlahos, O. Lu-Steffes, J. A. Alonso, D. Morgan, and J. Booske, "Demonstration of low work function perovskite $SrVO_3$ using thermionic electron

emission," *Adv. Funct. Mater.* 32 (41), 2203703 (2022). https://doi.org/10.1002/adfm.202203703

[20] R. Xu, L. Min, Z. Qi, X. Zhang, J. Jian, Y. Ji, F. Qian, J. Fan, C. Kan, H. Wang, W. Tian, L. Li, W. Li, and H. Yang, "Perovskite Transparent Conducting Oxide for the Design of a Transparent, Flexible, and Self-Powered Perovskite Photodetector," *ACS Appl. Mater. Interfaces* 12, 16462-16468 (2020). https://doi.org/10.1021/acsami.0c01298

[21] K. K. Ghose, Yun Liu, and T. J. Frankcombe, "High-temperature reduction thermochemistry of $SrVO_{3-\delta}$," *J. Phys. Energy* 6, 015007 (2024). https://doi.org/10.1088/2515-7655/ad0b8a

[22] M. S. Sheikh, R. Jacobs, D. Morgan, and J. Booske, "Time-dependence of $SrVO_3$ thermionic electron emission properties," *J. Appl. Phys.* 135, 055104 (2024) https://doi.org/10.1063/5.0186012

[23] D. Chen, R. Jacobs, J. Petillo, V. Vlahos, K. L. Jensen, D. Morgan, and J. Booske, "Physics-Based Model for Nonuniform Thermionic Electron Emission from Polycrystalline Cathodes," *Phys. Rev. Applied* 18, 054010 (2022). https://doi.org/10.1103/PhysRevApplied.18.054010

[24] X. Liu, Q. Zhou, T. L. Maxwell, B. K. Vancil, M. J. Beck, and T. J. Balk, "Scandate cathode surface characterization: Emission testing, elemental analysis and morphological evaluation," *Mater. Charact.* 148, 188-200 (2019). https://doi.org/10.1016/j.matchar.2018.12.013

[25] D. Chen, R. Jacobs, D. Morgan, and J. Booske, "Physical Factors Governing the Shape of the Miram Curve Knee in Thermionic Emission," *IEEE Trans. Electron Devices* 70, 1219-1225 (2023). https://doi.org/10.1109/TED.2023.3239058.

[26] D. Chen, R. Jacobs, D. Morgan, and J. Booske, "Impact of Nonuniform Thermionic Emission on the Transition Behavior Between Temperature-and Space-Charge-Limited Emission," *IEEE Trans. Electron Devices*, 68, 3576-3581 (2021). https://doi.org/10.1109/TED.2021.3079876.

[27] J. M. Lafferty, "Boride Cathodes," *J. Appl. Phys.* 22, 299–309 (1951). https://doi.org/10.1063/1.1699946

[28] D. M. Kirkwood, S. J. Gross, T. J. Balk, M. J. Beck, J. Booske, D. Busbaher, R. Jacobs, M. E. Kordesch, B. Mitsdarffer, D. Morgan, W. D. Palmer, B. Vancil, and J. E. Yater, "Frontiers in thermionic cathode research," *IEEE Trans. Electron Devices* 65 (6), 2061-2071 (2018). https://doi.org/10.1109/TED.2018.2804484

[29] T. Berry, S. Bernier, G. Auffermann, T. M. McQueen, and W. A. Phelan, "Laser floating zone growth of $SrVO_3$ single crystals," *Journal of Crystal Growth* 583, 126518 (2022). https://doi.org/10.1016/j.jcrysgro.2022.126518

[30] C. Wan, J. M. Vaughn, J. T. Sadowski, and M. E. Kordesch, "Scandium oxide coated polycrystalline tungsten studied using emission microscopy and photoelectron spectroscopy," *Ultramicroscopy* 119, 106-110 (2012). https://doi.org/10.1016/j.ultramic.2011.10.001

[31] D. Cahen, and A. Kahn, "Electron Energetics at Surfaces and Interfaces: Concepts and Experiments," *Adv. Mater.* 15, 271-277 (2003). https://doi.org/10.1002/adma.200390065

[32] M. S. Sheikh, L. Lin, R. Jacobs, M. E. Kordesch, J. T. Sadowski, M. Charpentier, D. Morgan, J. Booske, Raw data for the manuscript entitled "Direct evidence of low work function on $SrVO_3$ cathode using thermionic electron emission microscopy and high-field ultraviolet photoemission spectroscopy." figshare Dataset (2024). https://doi.org/10.6084/m9.figshare.25202624

[33] W. Telieps, E. Bauer, "An analytical reflection and emission UHV surface electron microscope," *Ultramicroscopy* 17, 57–65 (1985).